\pdfoutput=1

\documentclass[11pt,twoside,a4paper,cmspaper,final,collab]{cms-tdr}
\def\svnVersion{460517P}\def\cmsCernNoTag{CERN-EP-2018-007}\def\cmsCernDate{\today}\def\cmsMessage{Published in Physical Review Letters as \href{http://dx.doi.org/10.1103/PhysRevLett.120.201801}{\doi{10.1103/PhysRevLett.120.201801.}}}

\begin{document}\cmsNoteHeader{EXO-16-057}

\hyphenation{had-ron-i-za-tion}
\hyphenation{cal-or-i-me-ter}
\hyphenation{de-vices}
\RCS$HeadURL: svn+ssh://svn.cern.ch/reps/tdr2/papers/EXO-16-057/trunk/EXO-16-057.tex $
\RCS$Id: EXO-16-057.tex 451900 2018-03-20 21:12:35Z glandsbe $

\newlength\cmsFigWidth
\ifthenelse{\boolean{cms@external}}{\setlength\cmsFigWidth{0.85\columnwidth}}{\setlength\cmsFigWidth{0.4\textwidth}}
\ifthenelse{\boolean{cms@external}}{\providecommand{\cmsLeft}{top\xspace}}{\providecommand{\cmsLeft}{left\xspace}}
\ifthenelse{\boolean{cms@external}}{\providecommand{\cmsRight}{bottom\xspace}}{\providecommand{\cmsRight}{right\xspace}}
\ifthenelse{\boolean{cms@external}}{\providecommand{\NA}{\ensuremath{\cdots}\xspace}}{\providecommand{\NA}{\ensuremath{\text{---}}\xspace}}
\ifthenelse{\boolean{cms@external}}{\providecommand{\CL}{C.L.\xspace}}{\providecommand{\CL}{CL\xspace}}

\newcommand{\mjj}{\ensuremath{m_\text{jj}}\xspace}
\newcommand{\sA}{\ensuremath{\sigma\mathcal{A}}\xspace}
\newcommand{\gp}{\ensuremath{g'_\PQq}\xspace}
\newcommand{\gpd}{\ensuremath{g'_\PQd}\xspace}
\newcommand{\mZp}{\ensuremath{m_{\cPZpr}}\xspace}
\newcommand{\GZp}{\ensuremath{\Gamma_{\cPZpr}}\xspace}

\cmsNoteHeader{EXO-16-057}

\title{\texorpdfstring{Search for narrow resonances in the \PQb-tagged dijet mass spectrum in proton-proton collisions at $\sqrt{s} = 8\TeV$}  {Search for narrow resonances in the b-tagged dijet mass spectrum in proton-proton collisions at sqrt(s) = 8 TeV}}

\date{\today}

\abstract{
A search for narrow resonances decaying to bottom quark-antiquark pairs is presented, using a data sample of proton-proton collisions at $\sqrt{s} = 8\TeV$ corresponding to an integrated luminosity of 19.7\fbinv. The search is extended to masses lower than those reached in typical searches for resonances decaying into jet pairs at the LHC, by taking advantage of triggers that identify jets originating from bottom quarks. No significant excess of events is observed above the background predictions. Limits are set on the product of cross section and branching fraction to bottom quarks for spin 0, 1, and 2 resonances in the mass range of 325--1200\GeV. These results improve on the limits for resonances decaying into jet pairs in the 325--500\GeV mass range.
}

\hypersetup{
pdfauthor={CMS Collaboration},
pdftitle={Search for narrow resonances in the b-tagged dijet mass spectrum in proton-proton collisions at sqrt(s) = 8 TeV},
pdfsubject={CMS},
pdfkeywords={CMS, physics, exotica, narrow resonance searches}}

\maketitle

 Searches for new particles decaying to pairs of jets are pursued vigorously at hadron colliders, repeated at every new energy with ever increasing sensitivity in the quest for physics beyond the standard model (SM). Such ``dijet" final states have been explored in proton-antiproton collisions by the UA1~\cite{UA1} and UA2~\cite{UA2-1,UA2-2} Collaborations at the CERN S$\Pp\Pap$S, and at $\sqrt{s} = 1.8$ and 1.96\TeV by the CDF~\cite{CDF00,CDF0,CDF1,CDF2,CDF3,CDF4} and D0~\cite{D0-1,D0-2,D0-3} Collaborations at the Fermilab Tevatron, as well as  in proton-proton ($\Pp\Pp$) collisions at $\sqrt{s} = 7$, 8, and 13\TeV by the ATLAS~\cite{ATLAS1,ATLAS2,ATLAS3,ATLAS4,ATLAS5,ATLAS6,ATLAS7,ATLAS8,ATLAS9,ATLAS10} and CMS~\cite{CMS1,CMS2,CMS3,CMS4,CMS5,CMS6,CMS7,CMS8,CMS9,CMS10,CMS11,CMS12,CMS13} Collaborations at the CERN LHC.

The LHC dijet searches currently explore both the high-mass end of the spectrum, not previously accessible at lower-energy machines, and the low-mass range, aiming to gain sensitivity to much smaller couplings than those probed by earlier experiments. The latter searches are much more difficult because of the very large backgrounds, which result in overwhelming event rates that are beyond the typical trigger bandwidth of the ATLAS and CMS experiments. To address this challenge, several novel search strategies have been considered.

Recently, the CMS Collaboration introduced the idea of a trigger-level analysis,  which profits from the fact that the trigger acceptance rate can be increased significantly if the size of the event is kept small. Thus, it is possible to collect events at an increased rate using a specialized trigger-level data output, which keeps only minimal information about the event. The trigger-level analysis, also referred to as a ``scouting analysis" in CMS, enabled the mass reach of LHC dijet searches to be extended down to masses as low as 500\GeV~\cite{CMS11}.

Another way of lowering the mass reach of dijet searches is to use an initial-state radiation (ISR) jet or photon to trigger on an event and analyze the dijet system recoiling against the ISR object. Given that the ISR triggers typically require a rather high threshold for the transverse momentum (\pt) of the ISR object, for sufficiently light resonances the two jets from their decays may be merged and reconstructed as a single large-radius jet. The mass of such a jet, determined using the so-called jet substructure techniques, can then be used to search for new light resonances. Searches of this kind, recently pioneered by CMS~\cite{CMS-trijets1,CMS-trijets2}, for the first time can reach resonance masses as low as 50\GeV, \ie, well below the lowest previously probed mass of 140\GeV, achieved by the UA2 analyses~\cite{UA2-1,UA2-2}. A similar analysis has been very recently carried out also by ATLAS~\cite{ATLAS-ISR}.

Yet another strategy of extending the reach to lower masses, pursued in this Letter, is to look for resonances decaying into jets originating from the fragmentation of \PQb quarks. The dominant QCD background in the \bbbar final states is significantly reduced compared to that in generic dijet final states, allowing for lower trigger thresholds and increased search sensitivity, particularly for resonances decaying preferentially into third-generation particles.

Beyond the SM theories predict a variety of such resonances, \eg, \cPZpr{} resonances in top-assisted technicolor models~\cite{TC}, Kaluza--Klein excitations of the graviton in the Randall--Sundrum (RS) models~\cite{RS1,RS2} with SM particles allowed to propagate in the bulk space~\cite{RS3}, or additional scalar or pseudoscalar resonances with Yukawa-like couplings to quarks, as expected in the general class of two Higgs doublet models~\cite{2HDM} or models with spin-0 dark matter mediators~\cite{Buckley:2014fba,Haisch:2015ioa}. However, even for resonances not preferentially decaying to \bbbar final states, the sensitivity of a \PQb quark dijet search may rival that of the generic searches because of the drastically reduced backgrounds. Searches for new, massive resonances decaying to \bbbar final states have been explored for the first time by the CMS~\cite{CMS9}  and ATLAS~\cite{ATLAS9} Collaborations. Yet, these searches only relied on the standard jet triggers used in the generic dijet searches, and therefore the minimum mass probed was as high as 1100 (ATLAS) or 1200  (CMS) \GeV. It is of particular importance to extend the mass reach of these searches below 1000\GeV, for which the existing limits are still rather weak. Moreover, the \bbbar channel is particularly important for resonances with enhanced couplings to third-generation particles and with masses below the \ttbar threshold of about 350\GeV.

The above three strategies are complementary to each other, as they vary in sensitivity to different production and decay mechanisms of new, light resonances. This Letter presents the first search for \bbbar resonances with masses as low as 325\GeV, \ie, below the \ttbar threshold, using dedicated triggers requiring the presence of  \PQb quark jets. The results improve upon the sensitivity of existing generic dijet searches to models predicting such resonances. The results are interpreted in the context of a spin-0 resonance, spin-1 \cPZpr{} boson, and spin-2 RS graviton, whose intrinsic widths are small compared to the experimental resolution.

The central feature of the CMS apparatus is a superconducting solenoid of 6\unit{m} internal diameter, providing a magnetic field of 3.8\unit{T}. Within the solenoid volume are a silicon pixel and strip tracker, a lead tungstate crystal electromagnetic calorimeter (ECAL), and a brass and scintillator hadron calorimeter (HCAL), each composed of a barrel and two endcap sections. Forward calorimeters extend the pseudorapidity ($\eta$) coverage provided by the barrel and endcap detectors. Muons are detected in gas-ionization chambers embedded in the steel flux-return yoke outside the solenoid. Events of interest are selected using a two-tiered trigger system~\cite{Khachatryan:2016bia}. The first level (L1), composed of custom hardware processors, uses information from the calorimeters and muon detectors to select events at a rate of around 100\unit{kHz} within a time interval of less than 4\mus. The second level, referred to as the high-level trigger (HLT), consists of a farm of processors running a version of the full event reconstruction software optimized for fast processing, and reduces the event rate to less than 1\unit{kHz} before data storage. A more detailed description of the CMS detector, together with a definition of the coordinate system used and the relevant kinematic variables, can be found in Ref.~\cite{Chatrchyan:2008zzk}.

The search is based on a sample of $\Pp\Pp$ collisions at a center-of-mass energy of 8\TeV collected with the CMS detector in 2012 and corresponding to an integrated luminosity of 19.7\fbinv.  The particle-flow (PF) event algorithm~\cite{ParticleFlow} aims to reconstruct and identify each individual particle with an optimized combination of information from the various elements of the CMS detector. The energy of photons is directly obtained from the ECAL measurement, corrected for zero-suppression effects. The energy of electrons is determined from a combination of the electron momentum, as determined by the tracker, the energy of the corresponding ECAL cluster, and the energy sum of all bremsstrahlung photons spatially compatible with originating from the electron track. The energy of muons is obtained from the curvature of the corresponding track. The energy of charged hadrons is determined from a combination of the momentum measured in the tracker and the matching ECAL and HCAL energy deposits, corrected for zero-suppression effects and for the response function of the calorimeters to hadronic showers. Finally, the energy of neutral hadrons is obtained from the corresponding corrected ECAL and HCAL energy. The missing transverse momentum, $\ptmiss$, is defined as the magnitude of the vectorial sum of transverse momenta of all PF candidates within the event.

Events are required to have at least one reconstructed collision vertex within 24 (2)\cm of the mean $\Pp\Pp$ interaction position
along the beam axis (in the plane transverse to the beams). The vertex with the highest sum of $\pt^2$ of all the associated tracks is taken to be the primary vertex in the event.

For each event, hadronic jets are clustered from both charged and neutral PF candidates using the infrared- and collinear-safe anti-\kt algorithm~\cite{Cacciari:2008gp} with a distance parameter of 0.5, as implemented in the \textsc{FastJet} package~\cite{Cacciari:2011ma}. Only those charged PF candidates originating from the primary vertex are included in the clustering. Corrections based on the jet area~\cite{Cacciari:2007fd} are applied to remove the energy contribution of neutral hadrons from additional $\Pp\Pp$ interactions within the same or nearby bunch crossings (pileup). The jet momentum is determined as the vectorial sum of all particle momenta in the jet, and is found in simulation to be within 5 to 10\% of the true generator-level jet momentum, over the whole \pt spectrum and detector acceptance considered in the analysis. Jet energy corrections are derived from the simulation, and are confirmed by in situ measurements of the energy balance of dijet, multijet, $\Pgg$+jet, and leptonically decaying $\PZ$+jet events~\cite{Chatrchyan:2011ds,Khachatryan:2016kdb}. Jet energy corrections are further propagated to \ptmiss. Additional selection criteria are applied to each event to remove spurious jet-like features originating from isolated noise patterns in certain HCAL regions~\cite{CMS-PAS-JME-10-003}. The jet energy resolution is typically 15\% at 10\GeV, 8\% at 100\GeV, and 4\% at 1\TeV.

Jets originating from \PQb quarks are identified using the combined secondary vertex (CSV) algorithm~\cite{Chatrchyan:2012jua,CMS-PAS-BTV-13-001}, which takes as inputs the impact parameters of jet constituents and secondary vertices reconstructed within the jet~\cite{Chatrchyan:2014fea}. We use the ``tight" (``medium") working point of the \PQb tagging algorithm, which corresponds to approximately 50 (70)\% b jet tagging efficiency and 0.1--0.2\% (1--2\%) light-quark or gluon jet mistag rate for jets with $\pt < 300\GeV$. The \PQb tagging efficiency in the simulation is corrected to match the efficiency measured in data~\cite{CMS-PAS-BTV-13-001}.

Simulated Monte Carlo (MC) samples are primarily used to model signal hypotheses, as background predictions are obtained directly from data. We consider three models of narrow resonances. Scalar resonance and RS graviton signal samples are generated at leading order (LO) with \PYTHIA~8.212~\cite{Sjostrand:2007gs}, which also models the parton shower and hadronization processes, using the CUETP8M1 underlying event tune~\cite{Monash,CUETP8M1}. The \cPZpr{} boson samples are generated with \MGvATNLO 2.3.3~\cite{Alwall:2014hca}, with the parton shower and hadronization modeled with \PYTHIA~8.  The scalar (\cPZpr) boson model assumes gluon-gluon (quark-antiquark) production, while the RS graviton model includes both gluon-gluon and quark-antiquark production mechanisms; in all three cases, only decays to bottom quarks are simulated. This is a conservative choice, as in the flavor-universal case decays to charm quark-antiquark pair would also contribute to the signal acceptance. However, since the charm quark tagging efficiency by a dedicated \PQb tagging algorithm is relatively low, we ignore this potential increase in the signal acceptance, leading to a conservative estimate of signal sensitivity. (We estimated the effect to be only 3--4\% in terms of the signal yield for a \cPZpr{}  boson with universal coupling to quarks.) For all signal hypotheses, the intrinsic resonance width is negligible compared to the experimental mass resolution. The scalar resonance and RS graviton signal samples use the NNPDF3.0LO parton distribution functions (PDFs)~\cite{Ball:2014uwa}, while the \cPZpr{} boson samples are generated with the NNPDF2.3LO PDF set~\cite{Ball:2012cx}. Eight mass hypotheses are simulated between 325 and 1200\GeV for each of the three signal models.

The QCD multijet background samples are used to guide the analysis optimization and to study the performance of the \PQb tagging algorithm. The samples are generated at LO using \PYTHIA~6.424~\cite{Sjostrand:2006za} with the CTEQ6L1 PDFs~\cite{Pumplin:2002vw} and the underlying event tune Z2*~\cite{Z2star1,CUETP8M1}.
For all MC samples, the response of the CMS detector is simulated using \GEANTfour~\cite{GEANT4}, including the effects of pileup, obtained by superimposing additional minimum bias interactions on the hard scattering, with the multiplicity distribution matching that in data.

Online, events are selected using dedicated triggers that identify jets originating from \PQb quarks at the HLT. At L1, either one jet with $\abs{\eta}<5$ and $\pt > 128$\GeV or two jets with $\abs{\eta} < 1.74$ and $\pt > 56$\GeV are required.  At the HLT, the jets are reconstructed solely from energy deposits in the calorimeter towers and augmented with the tracking information within the jet cone. Two triggers with different requirements on jet \pt and geometrical acceptance are used, defining the low-mass (SR1) and high-mass (SR2) signal regions. For SR1, the trigger requires two jets with $\abs{\eta}<1.7$, with the leading and subleading (in \pt) jets having $\pt>80$ and 70\GeV, respectively. For SR2, the two jets are required to satisfy $\abs{\eta}<2.2$, with the leading (subleading) jet $\pt>160$ (125)\GeV. The HLT \PQb tagging algorithm requires that the ratio of the impact parameter to its uncertainty (including the uncertainty in the primary vertex position) is large for at least two tracks within the jet area~\cite{CMS-PAS-BTV-13-001}. At least two of the leading six jets in the event are required to satisfy the HLT \PQb tagging requirements. For signal events passing the rest of the event selection, the efficiency of the trigger \PQb tagging algorithm is approximately 18\% for SR1 and 49\% for SR2, as determined from combined studies based on collision data dominated by QCD multijet events, as well as on signal and QCD multijet background simulations. The trigger efficiency stays constant within the uncertainties as a function of the invariant mass of the two \cPqb\ -tagged jets, in the entire range used in the analysis.

\begin{figure}[tb]
	\centering
	\includegraphics[width=0.48\textwidth]{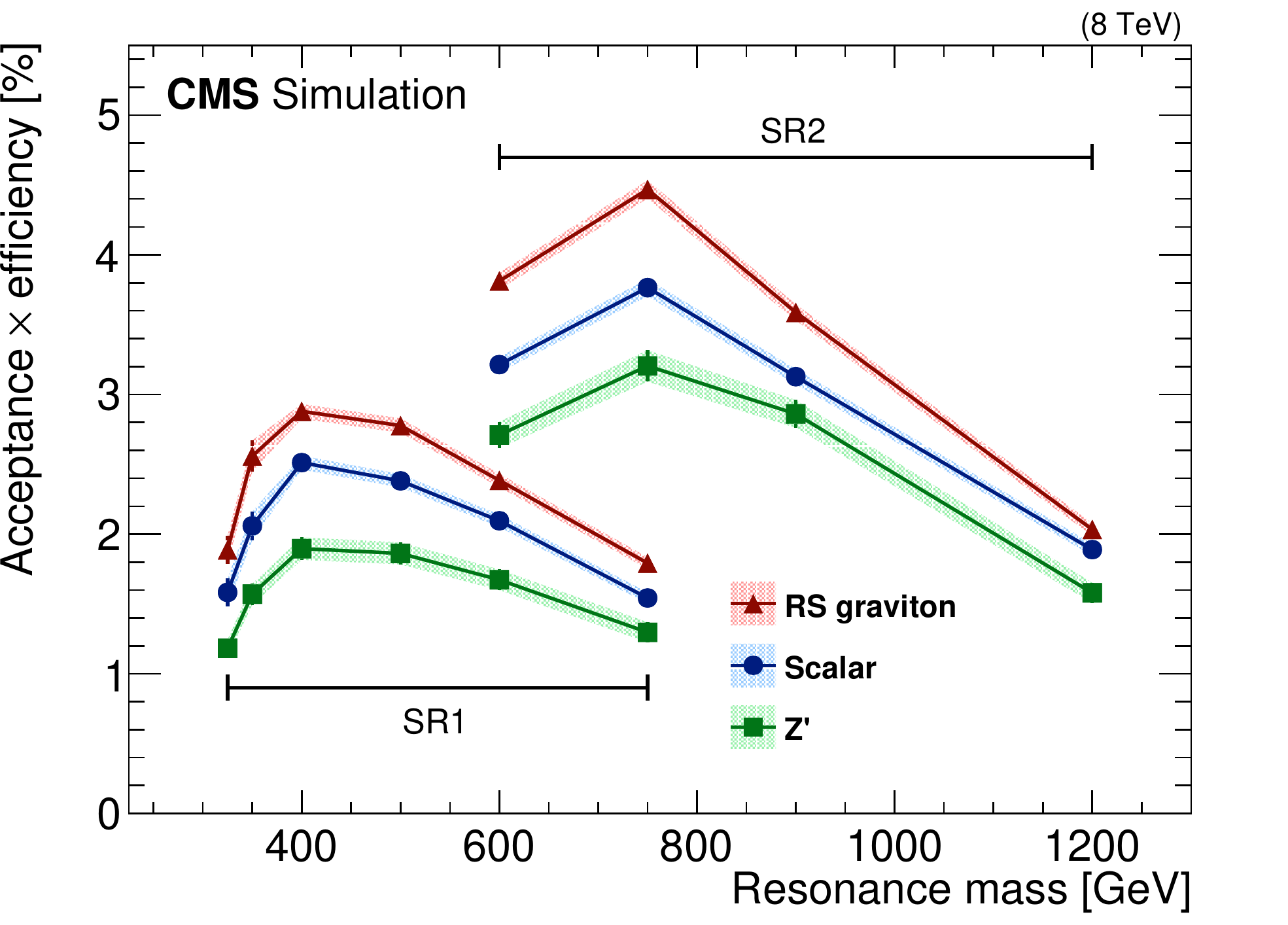}
	\caption{The products of acceptance and efficiency for simulated signal events in SR1 and SR2, separately for the scalar, \cPZpr, and RS graviton signal models. The shaded bands represent the statistical uncertainties.}
	\label{fig:signal-efficiency}
\end{figure}

Offline, jets built from PF candidates are used. Events are required to satisfy $\ptmiss/\sum\ET<0.5$, where $\sum\ET$ is the scalar sum of the transverse momenta of the PF candidates in the event. This requirement removes events with the energy of one of the jets significantly mismeasured, as well as events with large calorimeter noise inside a jet. The two leading jets form the dijet system. The jets must satisfy the same \pt and $\eta$ requirements as in the corresponding HLT trigger. The pseudorapidity difference between the two jets must be less than $1.3$. This requirement reduces the QCD multijet background considerably, while retaining high signal efficiency~\cite{ATLAS1,CMS1}. One of the two leading jets is required to pass the tight working point of the CSV algorithm, while the other must pass the medium CSV working point. Finally, the dijet invariant mass (\mjj) range is set to 296--1058\GeV for SR1, and 526--1607\GeV for SR2. These two search regions are used to probe signal masses in the range 325--700 and 700--1200\GeV, respectively, with the boundary chosen in the vicinity of the intersection of the expected limits in these two regions for all three resonance spins.

The product of acceptance and efficiency (\sA) for simulated signal events are shown in Fig.~\ref{fig:signal-efficiency}. For SR1 (SR2), these range from 1.2 to  2.9\% (1.6 to 4.5\%), with small differences between models due to differences in the geometrical acceptance, defined by the rapidity requirements on the two leading jets. At high masses (above 750\GeV), \sA drops because of the reduced \PQb tagging efficiency for high-\pt jets.

\begin{figure*}[t]
	\centering
\includegraphics[width=0.48\textwidth]{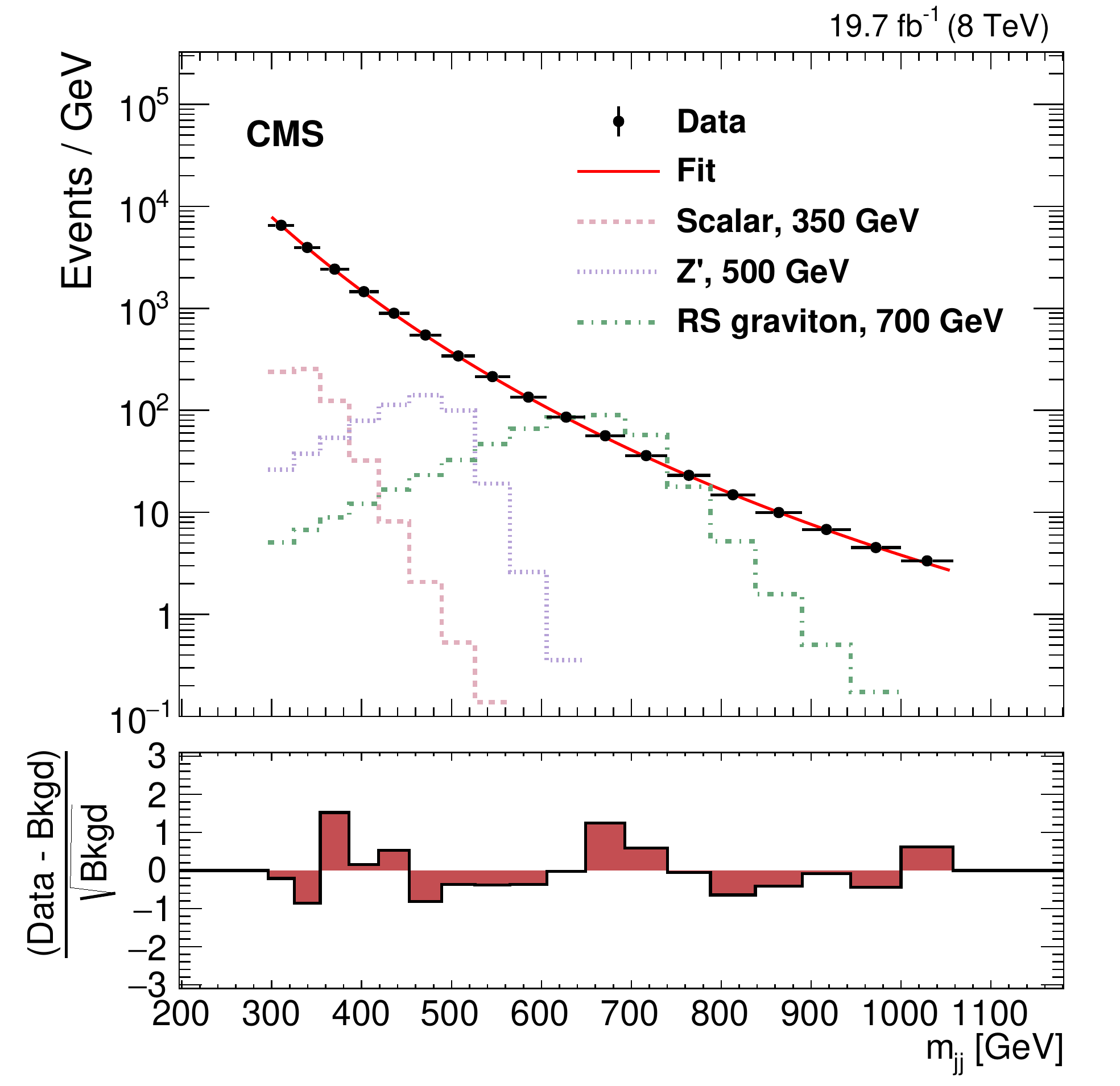}
\includegraphics[width=0.48\textwidth]{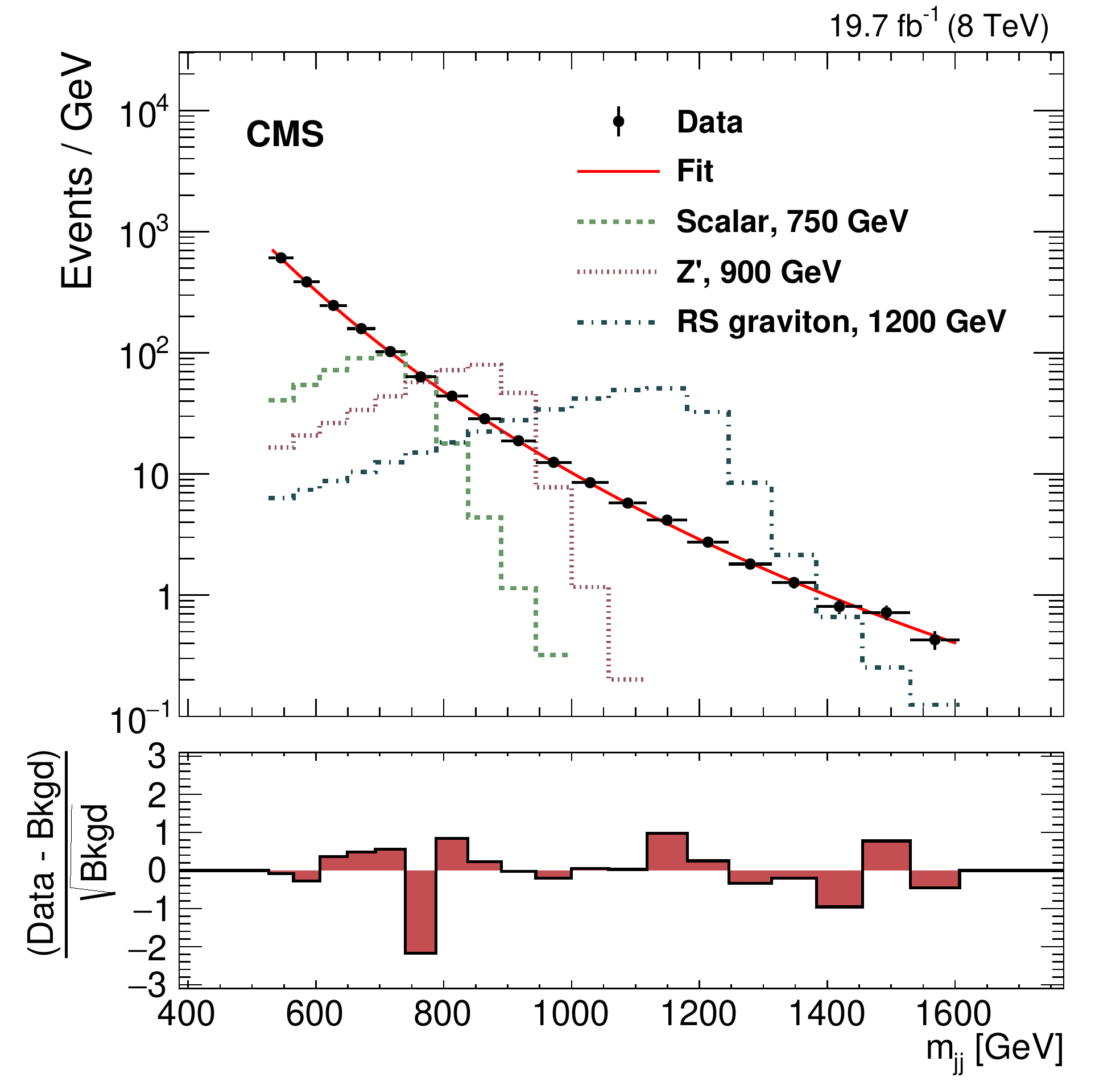} 	\caption{The dijet invariant mass distributions in SR1 (left) and SR2 (right), shown with the background prediction derived from a fit using an empirical function under the background-only hypothesis. Representative examples of signal distributions are also shown, each normalized to a visible cross section of 1\unit{pb}. The bottom panels show the difference between the data and the background estimate, divided by the statistical uncertainty in the estimated background.}
	\label{fig:mjj}
\end{figure*}

The background estimate is obtained from a binned (with 1\GeV bins), extended maximum likelihood~\cite{Barlow} fit to the \mjj spectrum in data using an empirically determined function. Several families of steeply falling functions commonly used in similar searches are considered, and the best fit function is chosen using an $F$-test~\cite{F-test} based on the $\chi^2$ per degree of freedom of the fit. The function chosen is: $\rd\sigma/{\rd}x = \epsilon_{\text{trig}}(x)p_0 (1-x)^{p_1}x^{-p_2-p_3\log(x)}$, where $x=\mjj/\sqrt{s}$, and $\epsilon_{\text{trig}}(x)$ is a sigmoid function describing the efficiency of the \pt requirements of the trigger. The parameters of the sigmoid function are determined in events collected with triggers requiring a single isolated muon, and are fixed in the background fit. The trigger turn-on effect is sizable only at the lower end of SR1, with the trigger inefficiency being 1.8\% for $\mjj=296\GeV$ and less than 0.1\% for $\mjj>380\GeV$. The \mjj distributions of the signal hypotheses are modeled using convolutions of a Gaussian and an exponential function~\cite{Bukin}. The signal shapes for masses between two adjacent simulated mass points are derived via a linear interpolation of the fit function parameters. The typical width of the Gaussian core of a signal resonance is 10--15\%, depending on the resonance spin and production mechanism, as well as on the resonance mass.

Extensive studies of a possible systematic bias from the choice of the functional form of the background estimate are performed with alternative fit functions, with or without signal injection. The shapes obtained from background-only fits to the data with the alternative functions are used to generate pseudo-data sets. Each pseudo-data set has a total number of events randomly drawn from a Poisson distribution with the mean equal to the yields observed in data. In the set of studies with signal injection, the pseudo-data sets are generated from a signal plus background model. In these studies, the injected signal cross section corresponds approximately to the expected 95\% confidence level (\CL) cross section limits discussed below. The generated \mjj spectra are then fitted with the sum of chosen background function and a signal model, and the signal cross section is extracted. Distributions of the difference between the fitted and injected signal cross sections divided by the fitted uncertainty are constructed, and their shapes are found to be consistent with a normal distribution with the mean within 0.5 of zero and the width consistent with unity. Thus, we conclude that any possible systematic bias from the choice of the functional form is small compared to the statistical uncertainty of the fit, and use the latter as the only uncertainty in the background prediction.

Figure~\ref{fig:mjj} shows the \mjj distributions in data in SR1 and SR2, fitted with the background-only hypothesis, together with representative examples of signal distributions normalized to a visible cross section, \ie, fiducial \sA of 1\unit{pb}. For presentation purposes, the data are binned with a bin width approximating the experimental dijet mass resolution.

Systematic uncertainties are assigned to the simulated signal to account for observed differences between simulation and data. Jet energy scale and resolution uncertainties of 1 and 10\%~\cite{Khachatryan:2016kdb} in the dijet invariant mass, respectively, are included as the uncertainties in the fitted signal parameters. The following four sources of uncertainty in the signal yield are considered. Scale factors are applied to account for mismodeling of the \PQb tagging efficiency in simulation, leading to a 5--15\%~\cite{Chatrchyan:2012jua,CMS-PAS-BTV-13-001} uncertainty, depending on the signal mass. An uncertainty of 10\% is assigned to the \PQb tagging efficiency in the HLT (as measured from data collected with unbiased prescaled triggers). An uncertainty of 2--5\% is assigned to account for the effect of the choice of PDFs on the signal acceptance, following the PDF4LHC prescription~\cite{PDF4LHC1,PDF4LHC2}. Finally, an uncertainty of 2.6\% is assigned to the integrated luminosity measurement~\cite{CMS-PAS-LUM-13-001}.

\begin{figure*}[htb]
	\centering
\includegraphics[width=0.32\textwidth]{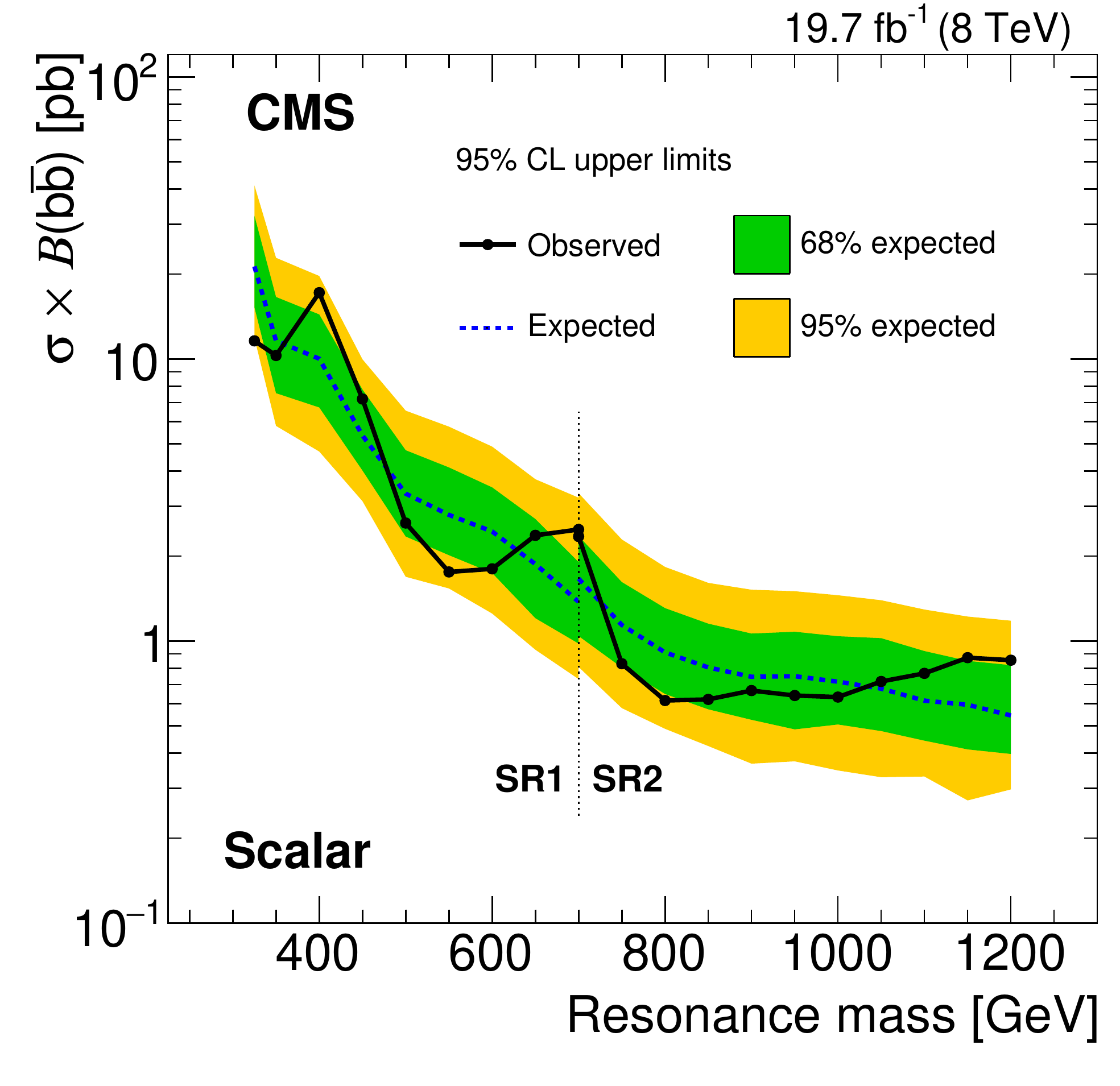}
\includegraphics[width=0.32\textwidth]{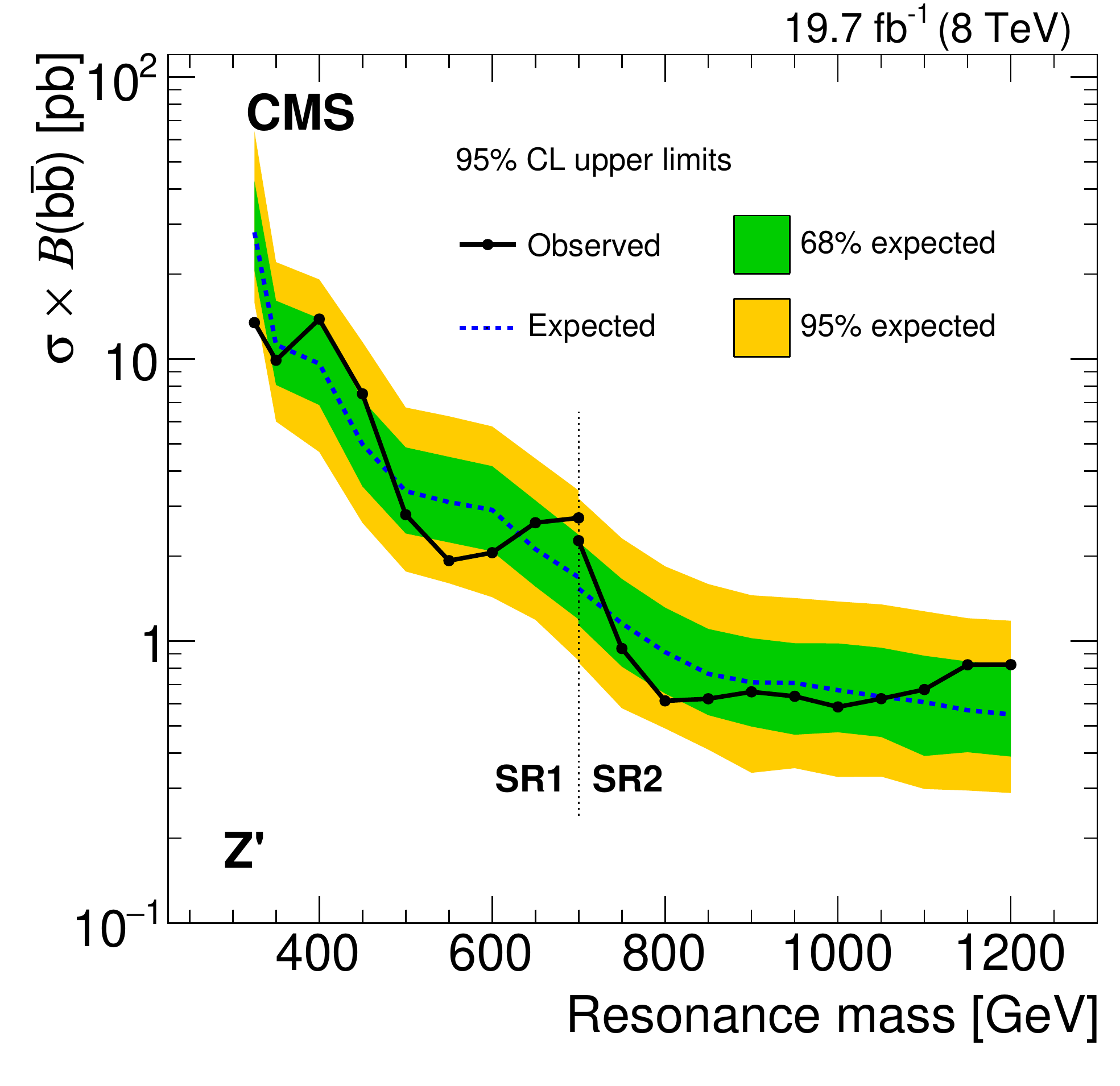}
\includegraphics[width=0.32\textwidth]{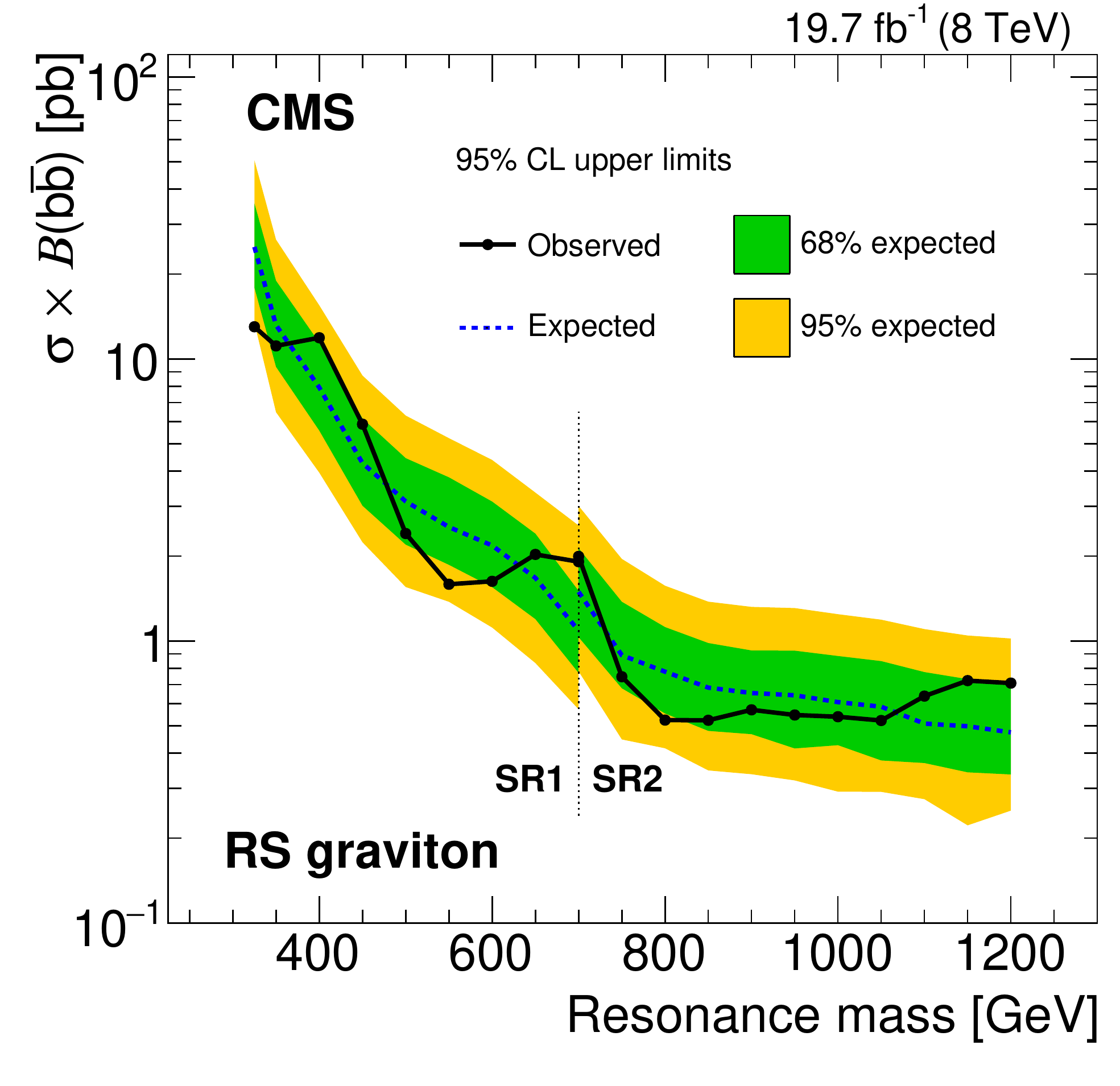}
	\caption{Observed and expected 95\% \CL upper limits on the product of cross section and branching fraction to bottom quark-antiquark pairs for a scalar resonance (left), \cPZpr{} boson (middle), and RS graviton (right) signal models, as functions of resonance mass. The discontinuity in the limits at 700\GeV is associated with a change in the acceptance from SR1 to SR2.}
	\label{fig:limits}
\end{figure*}
\begin{figure*}[htb]
	\centering
\includegraphics[width=0.43\textwidth]{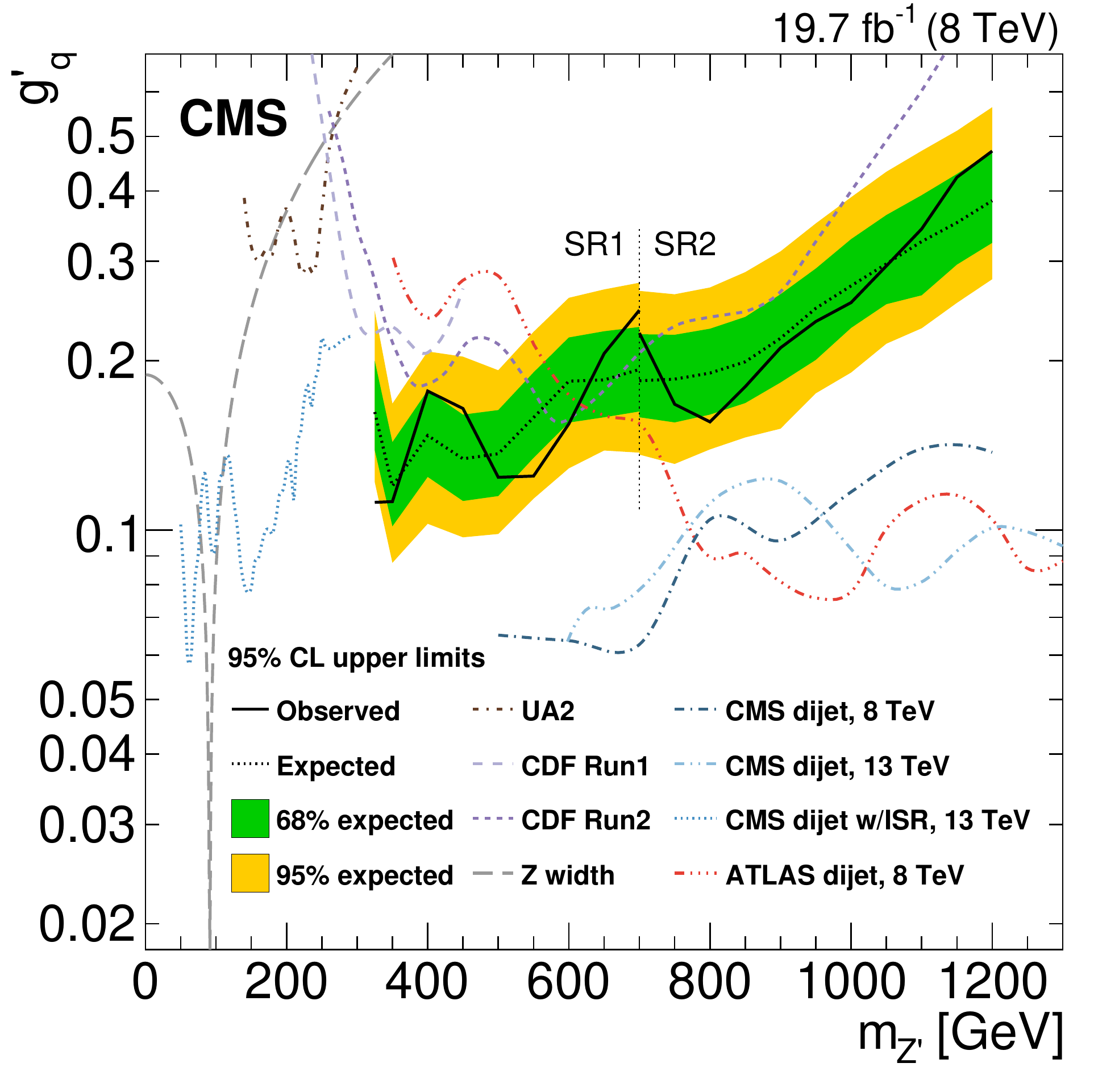}
\includegraphics[width=0.455\textwidth]{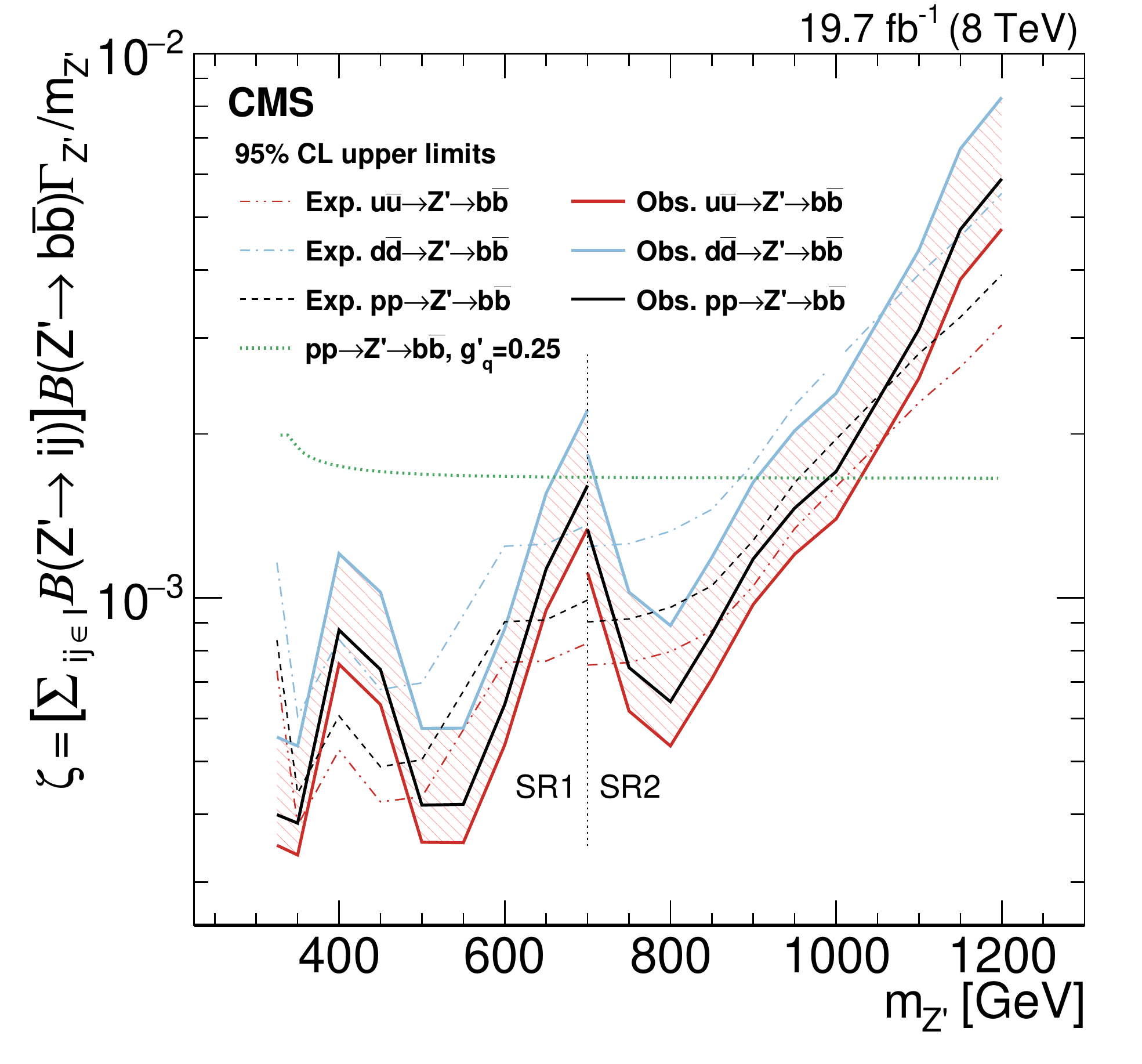}
	\caption{Left: The 95\% \CL upper limits (solid line)  on the universal coupling \gp between the leptophobic \cPZpr{} boson and quarks. Limits from other experiments~\protect\cite{UA2-1,CDF3,CDF4,ATLAS6} and earlier CMS analyses~\protect\cite{CMS11,CMS12,CMS-trijets2}, are also shown, along with an indirect constraint from the $\PZ$ boson width~\protect\cite{Dobrescu:2014fca}. Right: Expected (dashed lines) and observed (solid lines) 95\% \CL upper limits on the simplified model variable $\zeta$. The limits are shown for $\PQu\PAQu\to\cPZpr$ and $\PQd\PAQd\to\cPZpr$ individually, as well as for $\Pp\Pp\to\cPZpr$, assuming a universal quark coupling. The $\zeta$ values for the \cPZpr{} boson model with $\gp = 0.25$ are also shown. The hatched red band represents the envelope of limits for theoretical models that predict an $s$-channel production of a \cPZpr{} resonance with arbitrary couplings to up and down quarks. The discontinuity in the limits at 700\GeV is associated with a change in the acceptance from SR1 to SR2.}
	\label{fig:gq}
\end{figure*}

For each signal hypothesis, the dijet invariant mass spectrum is fit with a signal plus background hypothesis, where the parameters of the background function are freely floating. No significant excesses over the background-only hypothesis are observed. We set limits on the production of narrow resonances using the CL$_\mathrm{s}$ criterion~\cite{Junk:1999kv,Read:2002hq,ATL-PHYS-PUB-2011-011}, with an asymptotic approximation~\cite{Cowan:2010js} for the likelihood ratio used as a test statistic, and log-normal (Gaussian) constraints used to account for the systematic uncertainties in the signal and background yields (shapes). In Fig.~\ref{fig:limits}, the results are interpreted as upper limits at 95\% \CL on the product of cross section and branching fraction to bottom quark-antiquark pairs, $\mathcal{B}(\bbbar)$. The observed limits improve on the previously obtained limits on the \bbbar resonances for masses below 1.1\TeV, and extend below the \ttbar threshold, which is important to restrict the models with resonances coupled preferentially to the third-generation particles.

The limits on the \cPZpr{} boson model are further interpreted in the context of a simplified model of a leptophobic vector resonance with a universal coupling to quarks \gp that is related to the coupling of Ref.~\cite{Dobrescu:2013coa} by $\gp=g_{\mathrm{B}}/6$. The limits on \gp are shown in Fig.~\ref{fig:gq} (left), along with limits from other experiments~\cite{UA2-1,CDF3,CDF4,ATLAS6} and earlier CMS analyses~\cite{CMS11,CMS12,CMS-trijets2}. The current results improve on the existing limits in the \cPZpr{} mass range $325<\mZp<500\GeV$, where \gp values above 0.11--0.18 are excluded. We note that the narrow-width approximation used in setting cross section limits in this analysis that are further translated into \gp limits is valid only for \gp values $\lesssim$0.7. This upper limit corresponds to a resonance width of about 25\% of its mass, \ie, comparable with the instrumental resolution. Consequently, we truncate the $y$ axis of Fig.~\ref{fig:gq} (left) at this value of the coupling.

Following the method described in Ref.~\cite{Chivukula:2016hvp}, the limits on the \cPZpr{} boson model are further interpreted as limits on the variable $\zeta=[\sum_{ij\;\in\; I}\mathcal{B}(\cPZpr\to ij)]\mathcal{B}(\cPZpr\to \bbbar) \GZp/\mZp$, where \GZp is a width of the \cPZpr{} resonance, $\mathcal{B}$ is a branching fraction, and $I$ represents the set of production modes $ij \to \cPZpr$, with $i$ and $j$ being the corresponding partons. The $\zeta$ variable provides a model-independent description of the generic $s$-channel production of narrow-width resonances and can be used for a variety of theoretical interpretations of experimental limits on the production of such resonances decaying into various final states. The limits are shown in Fig.~\ref{fig:gq} (right) for the \cPZpr{} model with a universal quark coupling, as well as for up and down quark production modes individually. The limits are determined using the narrow-width approximation, which corresponds to a conservative interpretation~\cite{Chivukula:2017lyk}: for the \cPZpr{} boson model with $\gp=0.25$, the $\zeta$ limits computed with the resonance width taken into account are lower by 0.3 (4.7)\% at $\mZp=400$ (1200)\GeV. The $\zeta$ interpretation can be used, \eg, to convert the \gp limits in Fig.~\ref{fig:gq} to limits on the coupling \gpd for a \cPZpr{} boson model with coupling only to down-type quarks. Taking into account the different branching fractions and the widths of the two models, $\gpd=\gp[\zeta(\PQd\PAQd\to \cPZpr \to\bbbar)/\zeta(\Pp\Pp \to \cPZpr\to \bbbar)]^{1/2}$.

In summary, a search for new resonances decaying to bottom quark-antiquark pairs produced in 8\unit{TeV} proton-proton collisions has been presented. Using triggers that identify jets originating from bottom quarks, the search probes signal masses as low as 325\GeV. No statistically significant excesses above the background predictions are observed in the entire invariant mass range studied, 325--1200\GeV. Upper limits are set on the production cross section of scalar, vector, and tensor resonances. The limits are also interpreted in the context of a simplified model of a leptophobic \cPZpr{} boson with a universal coupling \gp to quarks. Values of \gp above 0.11--0.18 are excluded for \cPZpr{} boson masses below 500\GeV, improving on the previous best limits in this mass range, which date back to the CDF experiment. The first experimental limits on the parameter $\zeta$ of a simplified $s$-channel resonance framework~\cite{Chivukula:2016hvp} have been obtained, making possible the reinterpretation of the limits in a variety of theoretical models corresponding to different resonance production and decay mechanisms.

\begin{acknowledgments}
 We congratulate our colleagues in the CERN accelerator departments for the excellent performance of the LHC and thank the technical and administrative staffs at CERN and at other CMS institutes for their contributions to the success of the CMS effort. In addition, we gratefully acknowledge the computing centers and personnel of the Worldwide LHC Computing Grid for delivering so effectively the computing infrastructure essential to our analyses. Finally, we acknowledge the enduring support for the construction and operation of the LHC and the CMS detector provided by the following funding agencies: BMWFW and FWF (Austria); FNRS and FWO (Belgium); CNPq, CAPES, FAPERJ, and FAPESP (Brazil); MES (Bulgaria); CERN; CAS, MoST, and NSFC (China); COLCIENCIAS (Colombia); MSES and CSF (Croatia); RPF (Cyprus); SENESCYT (Ecuador); MoER, ERC IUT, and ERDF (Estonia); Academy of Finland, MEC, and HIP (Finland); CEA and CNRS/IN2P3 (France); BMBF, DFG, and HGF (Germany); GSRT (Greece); OTKA and NIH (Hungary); DAE and DST (India); IPM (Iran); SFI (Ireland); INFN (Italy); MSIP and NRF (Republic of Korea); LAS (Lithuania); MOE and UM (Malaysia); BUAP, CINVESTAV, CONACYT, LNS, SEP, and UASLP-FAI (Mexico); MBIE (New Zealand); PAEC (Pakistan); MSHE and NSC (Poland); FCT (Portugal); JINR (Dubna); MON, RosAtom, RAS, RFBR and RAEP (Russia); MESTD (Serbia); SEIDI, CPAN, PCTI and FEDER (Spain); Swiss Funding Agencies (Switzerland); MST (Taipei); ThEPCenter, IPST, STAR, and NSTDA (Thailand); TUBITAK and TAEK (Turkey); NASU and SFFR (Ukraine); STFC (United Kingdom); DOE and NSF (USA).
\end{acknowledgments}

\bibliography{auto_generated}
\cleardoublepage \appendix\section{The CMS Collaboration \label{app:collab}}\begin{sloppypar}\hyphenpenalty=5000\widowpenalty=500\clubpenalty=5000\textbf{Yerevan Physics Institute,  Yerevan,  Armenia}\\*[0pt]
A.M.~Sirunyan, A.~Tumasyan
\vskip\cmsinstskip
\textbf{Institut f\"{u}r Hochenergiephysik,  Wien,  Austria}\\*[0pt]
W.~Adam, F.~Ambrogi, E.~Asilar, T.~Bergauer, J.~Brandstetter, E.~Brondolin, M.~Dragicevic, J.~Er\"{o}, A.~Escalante Del Valle, M.~Flechl, M.~Friedl, R.~Fr\"{u}hwirth\cmsAuthorMark{1}, V.M.~Ghete, J.~Grossmann, J.~Hrubec, M.~Jeitler\cmsAuthorMark{1}, A.~K\"{o}nig, N.~Krammer, I.~Kr\"{a}tschmer, D.~Liko, T.~Madlener, I.~Mikulec, E.~Pree, N.~Rad, H.~Rohringer, J.~Schieck\cmsAuthorMark{1}, R.~Sch\"{o}fbeck, M.~Spanring, D.~Spitzbart, A.~Taurok, W.~Waltenberger, J.~Wittmann, C.-E.~Wulz\cmsAuthorMark{1}, M.~Zarucki
% ...
\vskip\cmsinstskip
% ...
1:~~Also at Vienna University of Technology, Vienna, Austria\\
% ...

\end{sloppypar}
\end{document}